\documentclass[12pt]{article}

\usepackage{color}
\usepackage{tabularx}

\usepackage{dsfont}
\usepackage{times}

\usepackage[dvips]{graphicx}
\usepackage{ifthen}
\usepackage{enumerate}
\usepackage{amsmath}
\usepackage{amssymb}
\usepackage[mathscr]{eucal}
\usepackage[errorshow]{tracefnt}
\usepackage{amsmath}
\usepackage{slashed}
\usepackage{amssymb}
\usepackage{array}
\usepackage{amsthm}
\usepackage{eucal}
\usepackage{eufrak}
\usepackage{epsf}
\usepackage{epsfig}
\usepackage{psfrag}
\usepackage{multirow}
\usepackage[apple mac]{inputenc}

\def\+{{+\!\!\!+}}

\def\Tr{\rm Tr}

\newcommand{\tr}{\text{tr}\,}

\def\pmb#1{\setbox0=\hbox{#1}%
\kern.0em\copy0\kern-\wd0
\kern-.04em\copy0\kern-\wd0
\kern.08em\copy0\kern-\wd0
\kern-.04em\raise.0433em\box0 }         

\newcommand{\nc}{\newcommand}
\nc{\beq}{\begin{equation}}
\nc{\eeq}[1]{\label{#1}\end{equation}}
\nc{\ber}{\begin{eqnarray}}
\nc{\eer}[1]{\label{#1}\end{eqnarray}}
\nc{\pek}[1]{\cite{#1}}
\nc{\enr}[1]{(\ref{#1})}
\nc{\kal}[1]{{\cal{#1}}}
\nc{\dott}{\;\cdot\;}

\def\0 {\nonumber}

\begin{document}

\setcounter{page}{0}
\newcommand{\inv}[1]{{#1}^{-1}} 
\renewcommand{\theequation}{\thesection.\arabic{equation}}
\newcommand{\be}{\begin{equation}}
\newcommand{\ee}{\end{equation}}
\newcommand{\bea}{\begin{eqnarray}}
\newcommand{\eea}{\end{eqnarray}}
\newcommand{\re}[1]{(\ref{#1})}
\newcommand{\qv}{\quad ,}
\newcommand{\qp}{\quad .}

\def\qp{Q_+}
\def\qm{Q_-}
\def\qbp{\bar Q_+}
\def\qbm{\bar Q_-}
\def\sgh{\Sigma_{g,h}}

\begin{titlepage}
\begin{center}

\hfill SISSA 03/2011/FM-EP\\

\vskip .3in \noindent


{\Large \bf{ Vertices, Vortices }} \\

{\Large \bf{ \&}}\\

{\Large \bf{Interacting Surface Operators}}\\

\vskip .2in

{\bf Giulio Bonelli, Alessandro Tanzini and Jian Zhao}

\vskip .05in
{\em\small
International School of Advanced Studies (SISSA) \\via Bonomea 265, 34136 Trieste, Italy \\
and \\ INFN, Sezione di Trieste }

\vskip .5in
\end{center}
\begin{center} {\bf ABSTRACT }
\end{center}
\begin{quotation}\noindent
We show that the vortex moduli space in non-abelian supersymmetric ${\cal N}=(2,2)$
gauge theories on the two dimensional plane with adjoint and anti-fundamental matter
can be described as an holomorphic submanifold of the instanton moduli space in four dimensions.
The vortex partition functions for these theories are computed via equivariant localization.
We show that these coincide with the field theory limit of
the topological vertex on the strip with boundary conditions corresponding to
column diagrams. Moreover, we resum the field theory limit of the vertex partition functions in 
terms of generalized hypergeometric functions formulating their AGT dual description as
interacting surface operators of simple type. 
Analogously we resum the topological open string amplitudes
in terms of q-deformed generalized hypergeometric functions
proving that they satisfy appropriate finite difference equations.

\end{quotation}
\vfill
\eject

\end{titlepage}
\tableofcontents
\section{Introduction}

Supersymmetric gauge theories in four dimensions are the building blocks for the most promising attempts
to formulate an extension of the standard model of particles bridging in a coherent and natural way to a unified 
picture at higher energies. This makes the study of these theories of paramount importance for high energy phenomenology.

String theory is actually the natural framework for a unified geometric description of supersymmetric gauge theories 
via geometric engineering \cite{vafa}. In particular, BPS protected sectors of the gauge theory are then described and computed 
exactly by means of topological strings \cite{topos}.
String theory can indeed produce more than just perturbative gauge theories. 
Actually, via a keen control on their non-perturbative sectors, string theory naturally engineers non perturbative extensions
of four dimensional gauge theories. For example,
these can be realized by the M-theory approach of Witten \cite{four} and its subsequent extension \cite{N=2}.
This formalism led to the celebrated AGT correspondence \cite{AGT}, stating the equality between the Nekrasov partition function \cite{Nek}
of the ${\cal N}=2$ four dimensional quiver gauge theory and the Liouville conformal block \cite{BPZ} 
on a surface encoding the quiver structure of the former.

The issues we will discuss here have to do with the interplay between different incarnations of 
counting problems in gauge and string theory. More precisely,
in this paper we will compute a given set of quantities which admit different interpretations
depending on the point of view one takes.
These different perspectives can be listed as follows:
\begin{itemize}
\item decoupling limit of surface operators in ${\cal N}=2$ four dimensional supersymmetric gauge theories
\item equivariant partition function of the two dimensional gauge theory on the defect surface
\item Chern-Simons theory on a Lagrangian submanifold of the dual toric Calabi-Yau geometry
\item AGT-dual as Toda conformal blocks with suitable degenerate field insertions 
\end{itemize}

The first perspective can be obtained via a D-brane construction by
suspending $N$ D4-branes between two parallel NS5-branes
and then by extending $N_f=N$ D2-branes between the D4-branes and
an external parallel NS5'-brane (see Fig. 1) \cite{surface}.
By rescaling one of the initial
NS5-brane to infinity, one freezes the four dimensional gauge theory dynamics, letting the
system at a classical phase \cite{bucov}.

The second point of view corresponds to focus on the leftover dynamics on the D2-branes \cite{ht}.
Its vacua structure is characterized by vortex configurations whose partition function should be systematically
computed. We make a detailed analysis of the derivation of these results from instanton counting and compare with
the related studies by Nekrasov and Shatashvili \cite{NS}.

The third corner is the viewpoint of the topological string on the system via geometric engineering.
Indeed, the D2/D4/NS5 system can be recast as the topological vertex \cite{tv} on the strip with suitable
representations on the external legs \cite{Iqbal}.

Finally, the AGT dual of the four dimensional gauge theory computation is produced by representing the surface
operators in the gauge theory \cite{wittensurf} as degenerate fields insertions in the Toda $A_{N-1}$ theory
\cite{surface,KPW,bucov,also}.
As we will see, the insertion point coordinates get interpreted as open moduli or vortex counting parameters
and the non-abelian vortex partition function can be interpreted as multiple surface operators of simple type in interaction.

The structure of this paper goes as follows.
In Section 2 we compute the vortex partition functions for adjoint and anti-fundamental matter in supersymmetric ${\cal N}=(2,2)$
gauge theories on the two dimensional plane via equivariant localization.
In Section 3 we compute the topological vertex on the strip with boundary conditions corresponding to
column diagrams on a side and empty or transposed diagrams on the other and we show that the field theory limit
of the open topological string amplitudes is equal to the vortex partition functions.
In Section 4 we resum the field theory limit of the vertex partition functions in terms of generalized hypergeometric functions
and therefore recover an AGT dual description in terms of degenerate Toda conformal blocks.
Furthermore, we discuss analogous resummation formulas for the topological open string amplitudes
in terms of q-deformed generalized hypergeometric functions.
Section 5 contains some concluding remarks and several observations on open questions and possible further developments.

\section{Vortices}

In this section we analyze the moduli space of vortices for $U(N)$ gauge theories
with an adjoint hypermultiplet,
$N_f =N$ fundamental matter multiplets
and
$N_a=N$ multiplets in the antifundamental representation.
The moduli space for $N_a=0$ and without the adjoint hypermultiplet was analysed in \cite{ht} via a proper D-brane construction.
This, as displayed in Fig.1, is obtained by considering a set of $k$ parallel D2 branes of finite size in one dimension
suspended between a NS5-brane and $N$ (semi-)infinite D4-branes.

\begin{figure}
\begin{center}
\includegraphics[ width=0.5\textwidth]{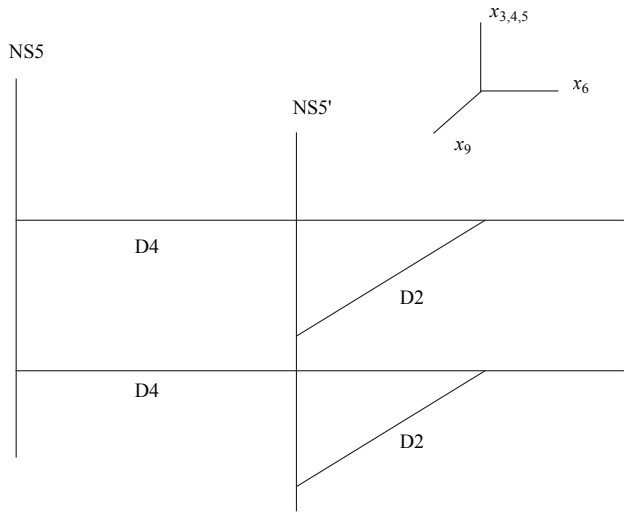}
\caption{Brane construction of surface operators}
\end{center}
\label{figure1}
\end{figure}

Interestingly, this moduli space
was found to be a holomorphic submanifold of the moduli space of instantons for an $U(N)$ ${\cal N}=2$ supersymmetric gauge theory
in four dimensions \cite{ht}. This was observed in the context of a brane construction of two-dimensional vortices in a four dimensional
gaug theory. The ADHM data are recovered via a double T-duality leading to a D0-D2 system. Let us notice that
an independent ADHM-like construction of the vortex moduli space was carried out in \cite{eto1} directly from field theory analysis
and shown in \cite{eto2} to be equivalent  
for $N_f=N$ to the D-brane construction of \cite{ht}\footnote{Further details are contained in \cite{shifman}.}
as far as the BPS state counting is concerned \cite{giappi}.
Here we will extend this analysis to the presence of adjoint and anti-fundamental matter and show that the relevant vortex moduli spaces
can be obtained as holomorphic submanifolds of the instanton moduli space of four dimensional ${\cal N}=2^*$ and ${\cal N}=2$ $N_f=N$ 
$U(N)$ gauge theories respectively.
Moreover, we will use equivariant localisation techniques to compute the relevant partition functions by vortex counting.

In order to study the moduli space in complete generality, we first consider the case ${\cal N}=(4,4)$, which we will
then reduce to $\mathcal{N}=(2,2)$ supersymmetry by turning on the relevant equivariant mass parameters. To this end
let us first recall the ${\cal N}=4$ ADHM construction of instantons following the notations of \cite{tanzinietc}. Indeed,
as we will show, the ${\cal N}=(4,4)$ vortex moduli space can be obtained as a holomorphic submanifold of this space. 
The ADHM data can be extracted from the low-energy dynamics of a system of $N$ $D3$-branes
and $k$ $D(-1)$-branes in flat space. In particular, the matrix model action for the $k$ $D(-1)$ branes contains
five complex fields $B_{\ell},\phi \in {\rm End}(V)$, $V=\mathbb{C}^k$ with $\ell=1,\ldots,4$ in the adjoint representation of $U(k)$ describing
the positions of the $k$ D(-1)-instantons in ten-dimensional space.
In addition open strings stretching
between D(-1)-D3 branes provide two complex moduli $I,J$ in the
$(\bar{k},N)$ and $(\bar{N},k)$ bifundamental representations respectively
of $U(k)\times U(N)$, that is $I\in {\rm Hom}(W,V)$ and $J\in {\rm Hom}(V,W)$ with $W=\mathbb{C}^N$.
The ADHM constraints can be read as D and F-term equations
of the matrix model action
\bea\label{Dterm}
&&[B_\ell,B_\ell^\dagger]+II^\dagger-J^\dagger J = \zeta
 \ \ , \nonumber\\
&&[B_1,B_2]+[B_3^\dagger, B_4^\dagger]+IJ =0 \ \ , \nonumber\\
&&[B_1,B_3]-[B_2^\dagger, B_4^\dagger] = 0 \ \ , \nonumber \\
&&[B_1,B_4]+[B_2^\dagger, B_3^\dagger] =0 \ \ ,
\eea
together with
\bea\label{Fterm}
&& B_3 I-B_4^\dagger J^\dagger = 0 \nonumber\\
&& B_4 I+B_3^\dagger J^\dagger = 0  \ \ .
\eea
The ${\cal N}=4$ instanton moduli space arises as a hyperkahler quotient with respect to a $U(k)$
group action with the above momentum maps (\ref{Dterm}) and (\ref{Fterm}).
We can obtain the vortex moduli space for the ${\cal N}=(4,4)$ theory in two dimensions by applying to the ADHM data
(\ref{Dterm}), (\ref{Fterm})
the same procedure developed in \cite{ht}, namely by
considering the Killing vector field
rotating the instantons in a plane and setting to zero the associated Hamiltonian. The vortices correspond then to instanton
configurations which are invariant under the selected rotation group. To be explicit, let us consider the following
$U(1)$ action
on the ADHM data
\bea
&&(B_1,B_2,B_3,B_4)\to (B_1, {\rm e}^{i\theta} B_2, B_3, {\rm e}^{-i\theta} B_4) \nonumber \\
&&(I,J)\to (I,{\rm e}^{i\theta} J)
\label{u1}
\eea
This is a Hamiltonian action with generating vector field
\be
\xi = {\rm Tr} \left(B_2\partial/\partial B_2 - B_4\partial/\partial B_4 + J\partial/\partial J - h.c.\right)
\label{exi}\ee
and Hamiltonian
\be
H = {\rm Tr} (B_2 B_2^\dagger + B_4 B_4^\dagger + JJ^\dagger ) \ \ .
\label{h}
\ee
Indeed we have
\be
i_{\xi}\omega^{(1,1)} =  d H
\ee
with the Kahler form
\be
\omega^{(1,1)}= dB_\ell\wedge dB^\dagger_\ell
+dJ^\dagger\wedge dJ +dI\wedge dI^\dagger.
\ee
By restricting the ${\cal N}=4$ ADHM data to the zero locus of the Hamiltonian (\ref{h})
we get a holomorphic submanifold described by the data $(B_1,B_3=\Phi)$ and $I$ subject
to the constraints
\bea\label{red-Dterm}
&&\left[B_1,B_1^{\dagger }\right]+\left[\Phi ,\Phi ^{\dagger
}\right]+I I^{\dagger }=\zeta \nonumber\\
&& [B_1,\Phi ]=0
\eea
together with the stability condition $\Phi  I=0$. The above data describe the moduli
space of $k$ vortices for $U(N)$ ${\cal N}=(4,4)$ gauge theory in two dimensions
as a Kahler quotient with $U(k)$ group action.
Indeed, (\ref{red-Dterm}) are the D-term equations for a supersymmetric euclidean D0-D2 system,
whose lagrangian can be obtained from the reduction of the ${\cal N}=2$
gauge theory in four dimensions with $N_f=N$ fundamentals. Its bosonic part reads
\bea\nonumber
{\cal L}&=&
\Tr\Bigl[
\frac{1}{2}[\Phi,\Phi^\dagger]^2 + \frac{1}{2}\left([B_1,B_1^\dagger]+II^\dagger-\zeta{\bf 1}\right)^2
+ \left\{\Phi,\Phi^\dagger\right\}II^\dagger + |\left[B_1,\Phi\right]|^2+
|\left[B_1,\Phi^\dagger\right]|^2
\\&&
+\frac{1}{2}[\varphi,\varphi^\dagger]^2
+|[\varphi,\Phi]|^2 +|[\varphi^\dagger,\Phi]|^2
+|[\varphi,B_1]|^2 + |[\varphi^\dagger,B_1]|^2
+\{\varphi,\varphi^\dagger\}II^\dagger
\Bigr]
\label{dt}\eea
where $\varphi$ is the complex scalar coming from the reduction of the four dimensional vector field,
and $\Phi$ is the complex scalar of the four-dimensional gauge theory.
The first line of (\ref{dt}), that is the $\varphi$ independent part of the potential, can be rewritten as
\bea\nonumber
&&\Tr\left[\frac{1}{2}[\Phi,\Phi^\dagger]^2 + \frac{1}{2}\left([B_1,B_1^\dagger]+II^\dagger-\zeta{\bf 1}\right)^2
+ \left\{\Phi,\Phi^\dagger\right\}II^\dagger + |\left[B_1,\Phi\right]|^2+
|\left[B_1,\Phi^\dagger\right]|^2\right]=\\
&&=\Tr\left[
\frac{1}{2}\left([B_1,B_1^\dagger]+[\Phi,\Phi^\dagger]+II^\dagger-\zeta{\bf 1}\right)^2
+2 \Phi II^\dagger\Phi^\dagger + 2 |\left[B_1,\Phi\right]|^2\right]
\label{jhgjh}\eea
while the second line of (\ref{dt}) contains the equivariant action on the fields generated by $\varphi$.

The D-term equations of (\ref{jhgjh}) correspond to the reduced ${\cal N}=4$ ADHM equations (\ref{red-Dterm}).

The vortex moduli space in presence of additional $N$ anti-fundamental matter multiplets can be obtained
with the same method by extending the above construction with anti-fundamental hypermultiplets with masses
$m_f$, $f=1,\ldots,N$, in the original four dimensional theory.
These contribute by giving extra fermion zero modes $\lambda_f$
with equivariant action $\varphi\cdot\lambda_f+ m_f\lambda_f$. These mass terms break to ${\cal N}=(2,2)$ supersymmetry.
We will now apply localization formulae in order to compute the vortex partition function.

\subsection{Counting vortices}

In this subsection we perform the computation of the non-abelian vortex partition function
via localization methods. Let us start with the case of the adjoint matter by computing
the fixed points
in the vortex moduli space under the torus action $T=T_{Cartan}\times T_{\hbar}\times T_m$, where
$T_{Cartan}=U(1)^{N}$ is the Cartan subgroup of the colour 
group\footnote{Notice that the colour group is identified with the flavour group in the two dimensional
theory after ungaugung and therefore the Cartan parameters become the mass parameters for the fundamental 
multiplets.}, $T_\hbar$ is the
lift to the vortices moduli space of the spatial rotation in $\mathbb{R}^2$
\be
\left(B_1,\Phi,I\right)\to \left(e^{i\hbar} B_1, \Phi, I\right)
\label{torus}\ee
and $T_m$ the $U(1)_R$ symmetry
\be
\left(B_1,\Phi,I\right)\to \left( B_1, e^{im} \Phi, I\right)
\label{torusm}\ee
where $m$ is the mass parameter of the four dimensional adjoint hypermultiplet breaking ${\cal N}=4$
to ${\cal N}=2^*$, which will become the mass of the adjoint scalar in two dimensions after the reduction.
We observe that the vortex action (\ref{jhgjh}) can be obtained from the ${\cal N}=2^*$ action upon reduction
under the Hamiltonian symplectomorphism generated by (\ref{exi}).

The classification of the fixed points proceeds in a way very similar to the instanton case, except that
now, since only the $B_1$ variable is involved, these are labeled by column diagrams
$\left\{1^{k_l}\right\}$ only, where $l=1\ldots,N$ and $\sum_l k_l=k$ is the total vortex number \footnote{See also 
the very recent paper \cite{Yoshida} for a similar computation.} where $k_l=\frac{1}{2\pi}\int_{\mathbb{C}} \tr(F\tau_l)$,
$\ \tau_l$ being the generator of the $l$-th Cartan subgroup.
In order to compute the determinants weighting the enumeration of fixed points in the localization formula,
we evaluate the equivariant character on the tangent space around the fixed points which provides the relevant eigenvalues.

The total equivariant character can be computed to be
\be
\tilde\chi=V^*\otimes V \left(T_\hbar + T_m^{-1}-1-T_m^{-1}T_\hbar\right) + V^*\otimes W \left(1- T_m^{-1}\right)
=\left(1-T_m^{-1}\right)\chi,
\label{eqchar}\ee
where the reduced character $\chi$ is given by
\be
\chi= V^*\otimes V\left(T_\hbar-1\right)+W^*\otimes V
\label{redchar}\ee

By exploiting the weight decomposition of the vector spaces
\be
V=\sum_{l=1}^N \sum_{i=1}^{k_l} T_{a_l}T_\hbar^{i-1},
\quad
W=\sum_{l=1}^N T_{a_l},
\label{wd}\ee
one easily computes the reduced character to be
\be
\chi=\sum_{l,m=1}^N \sum_{i=1}^{k_l}T_{a_{lm}}T_\hbar^{-k_m+i-1}.
\label{redchev}\ee

From (\ref{eqchar}), (\ref{redchar}) and (\ref{redchev}) we get
the determinant factor associated to a specific partition
${\bf k}=(k_1,\ldots,k_N)$
\be
Z^{adj}_{{\bf k}}=
\prod_{l,m}\prod_{i=1}^{k_l}
\frac{a_{lm}+\left(-k_m+i-1\right)\hbar-m}{a_{lm}+\left(-k_m+i-1\right)\hbar}
\label{adjoint-vortex}\ee
which is the partition function in presence of an adjoint multiplet of mass $m$.
In the infinite mass limit this provides a derivation of the
partition function corresponding to the $N_f=N$ theory
\be
Z^{vect}_{{\bf k}}=
\prod_{l,m}\prod_{i=1}^{k_l}
\frac{1}{a_{lm}+\left(-k_m+i-1\right)\hbar} \ .
\label{vector-vortex}\ee
Notice that the $m\to 0$ limit of (\ref{adjoint-vortex}) reduces to one. This is 
the expected result since in this limit we are recovering an enhanced ${\cal N}=(4,4)$
supersymmetric theory, which therefore we prove to compute the Euler characteristic of the vortex
moduli space.

Computing the partition function of vortices in presence of $N$ anti-fundamentals
with arbitrary masses amounts to
shift the reduced character $\chi$ by a factor $\delta\chi=-T_m V$ (see \cite{tanzinietc}), where now
$T_m=\otimes_{f=1}^N T_{m_f}$ is the generator of the $U(1)^{N_f}$ subgroup in $U(N_f)$.
The direct computation then gives
\be
Z^{af}_{{\bf k}}=\frac{
\prod_l\prod_f\prod_{i=1}^{k_l} a_{l}+(i-1)\hbar+m_f}
{\prod_{l,m}\prod_{i=1}^{k_l}
a_{lm}+\left(-k_m+i-1\right)\hbar}
\label{anti-vortex}\ee
that coincides with the result obtained by different methods in \cite{shadchin}, up to a shift $m_f\to m_f+\hbar$.

The generating functions for the abelian case are very simple, namely
\be
{\cal Z}^{vect}_{U(1)}=
\sum _{k=0}^{\infty }Z_{U(1),\, k}^{vect} z^k
=\sum _{k=0}^{\infty }z^k \prod _{i=1}^k\frac{1}{i \hbar }=\exp \left(\frac{\mathit{z}}{\hbar }\right)
\ee
for the pure vector contribution, while in presence of adjoint and anti-fundamental one gets respectively
\begin{eqnarray}
Z^{adj}_{U(1)}&\text{=}&\sum _{k=0}^{\infty }Z_{U(1),\, k}^{adj}\mathit{z}^k=\sum _{k=0}^{\infty }\mathit{z}^k\prod _{i=1}^k\frac{i+\frac{m}{\hbar }}{i}=(1-\mathit{z})^{-\frac{(m+\hbar)}{\hbar }}  \\
Z^{af}_{U(1)}&\text{=}&\sum _{k=0}^{\infty }Z_{U(1),\, k}^{af}\mathit{z}^k=\sum _{k=0}^{\infty }\mathit{z}^k\prod _{i=1}^k\frac{\left(\frac{a+m}{\hbar }+i-1\right)}{-i}=(1+z)^{\frac{-(a+m)}{\hbar }} \nonumber
\end{eqnarray}
These results match the ones of \cite{bucov}.


\subsection{Vortices from Instantons}

It is worth remarking that the above vortex counting can be recovered directly 
from instanton counting by reducing to Young diagrams of column type and setting 
the sum of the two equivariant parameters to zero\footnote{An analogous reduction
was considered in \cite{russi} for the special partition $\{k_1,\ldots,k_N\}=\{k,0,\ldots,0\}$.}.

Let us recall that \cite{PF}
\be
\chi_{inst}=\sum_{l,m}\sum_{s\in Y_l} T_{a_{lm}}
\left(
T_1^{-{\tt l}_l(s)}T_2^{{\tt a}_m(s)+1}
+
T_1^{{\tt l}_l(s)+1}T_2^{-{\tt a}_m(s)}
\right)
\label{inst}\ee
where ${\tt a}(s)$ and ${\tt l}(s)$ are the ``arm'' and ``leg'' of the $s^{th}$
box in the corresponding Young diagram.
Restricting the above formula (\ref{inst}) to column diagrams, $Y_l=1^{k_l}$, setting $T_1T_2=1$ and 
denoting $T_2=T_\hbar$, we get
\be
\chi_{inst}^{red.}=\sum_{l,m}\sum_{i=1}^{k_l} T_{a_{lm}}
\left[
T_\hbar^{k_m(s)-i+1}
+
T_\hbar^{-k_m+i-1}
\right]
=
\chi+\bar\chi
\ee
where $\bar\chi$ is the vortex character (\ref{redchar}) computed upon reflecting $\hbar\to-\hbar$.

The condition $T_1T_2=1$ is implied by the symplectic reduction.
Actually, the corresponding torus acts on the constraint $[B_1,B_2]+IJ=0$ which is identically vanishing on the 
Lagrangian submanifold which identifies the vortex moduli space. Therefore 
this torus action is trivial on the Lagrangian submanigold
and the corresponding equivariant parameter does not appear in the vortex partition function.

Analogously, one can compute the fundamental and adjoint matter contributions.
For the adjoint this is straightforwardly obtained by shifting by the mass $m$
the formula for the vector multiplet, while
the contribution to the instanton character of one (anti-)fundamental of mass $m_f$ is
\cite{tanzinietc}
\be
\left[\delta\chi^{af}_{inst}\right]^{red.}=-T_{m_f} \sum_l\sum_{i=1}^{k_l} T_{a_l}T_\hbar^{i-1}=\delta\chi
\ee
From the relation among the reduced instanton and the vortex character one one gets a straightforward relation
among the associated partition functions. For example, for the case of matter in the adjoint representation,
one gets
\be
\left[Z_{{\tt k}}^{inst}\right]^{red.} ({\bf a},m, \hbar) = Z_{\tt k}^{adj}({\bf a},m,\hbar) Z_{\tt k}^{adj}({\bf a},-m,\hbar)  
\label{adj-red}
\ee
This alternative derivation, on the view of the A-model geometric engineering of Nekrasov partition function 
in \cite{Iqbal}, points to a relation with open topological string amplitudes on a strip where
the reduction from arbitrary Young diagrams to columns is induced by suitably restricting the boundary
conditions on the toric branes.

Analogous considerations, leading to the computation of two dimensional superpotentials via limits
of the instanton partition function, were presented in \cite{NS}.
We would like to underline that our scaling limit is different and that, as we will discuss at the beginning of next 
section, corresponds to a classical limit in four dimensional gauge theories.
Indeed, the Nekrasov-Shatashvili limit corresponds to sending $\epsilon_2\to0$ at fixed coupling, while, as shown 
in \cite{GL}, the vortex partition functions can be recovered in a scaling limit in which also the gauge coupling
is involved. This on one side confirms our interpretation of the vortex counting as a classical limit of the four 
dimensional gauge theory and moreover suggests that our result could represent a specific sector of the 
Nekrasov-Shatashvili's one.

\section{Vertices}

In this section we describe the topological open string counterpart of the vortex counting functions
by using the topological vertex formalism.
Notice that in the previous paragraph we have shown that vortex counting can be obtained from
instanton counting at $\epsilon_1+\epsilon_2=0$. This implies that its topological string counterpart
is obtained in term of unrefined topological vertex.
 
The vortex partition function is identified with the
classical limit $\Lambda\to 0$ of the four dimensional gauge theory surface operator evaluation \cite{bucov}.
In the brane construction, this limit is realized by scaling to infinity the extension of the D4-brane in the $x^6$
direction.
From the viewpoint of the toric geometry engineering of the
four dimensional ${\cal N}=2$ gauge theory, this limit corresponds
to send to infinity the ladders of the relevant toric diagram, leaving us with a pure strip geometry, see Fig.2.

\begin{figure}
\begin{center}
\includegraphics[ width=0.7\textwidth]{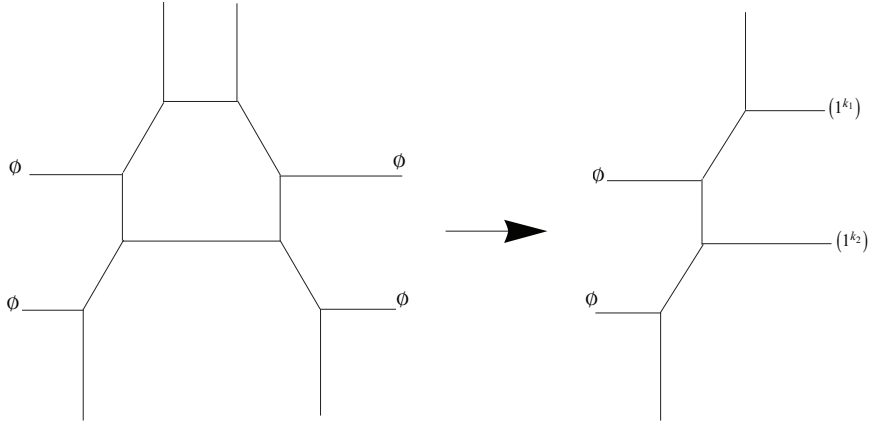}
\caption{Toric diagram engineering the 4D gauge theory and its classical limit to the strip}
\end{center}
\label{figure2}
\end{figure}

As we will show in the following,
the presence of the D2-branes is exactly taken into account by suitable boundary conditions on the topological vertex on the strip.
In particular, in the case of antifundamental matter one has to place on the internal legs column diagrams with lengths
$k_l$ $l=1,\ldots,N$, corresponding to the vortex number on each D2-brane ending on the $l$-th D4 brane, see Fig.3a. These correspond exactly to 
the column partitions of the total vortex number introduced in Sect.2.1.
The case of adjoint matter can be reproduced in the same setup by identifying the boundary conditions on the horizontal direction
of the toric diagram, see Fig. 3b.
This identification comes from the periodicity of the D-brane construction engineering the ${\cal N}=2^*$ theory.

Let us recall that the shape of the Young tableaux encoding the boundary conditions on the D-branes
corresponds to the choice of the representation of the gauge group of each inserted Wilson line.
Since the vortex vacua studied in the previous section
maximally break the gauge group as $SU(N)\to U(1)^{N-1}$, this has to be reproduced by the corresponding choice of
D-brane boundary conditions, namely by single column representations.

\begin{figure}
\begin{center}
\includegraphics[ width=0.7\textwidth]{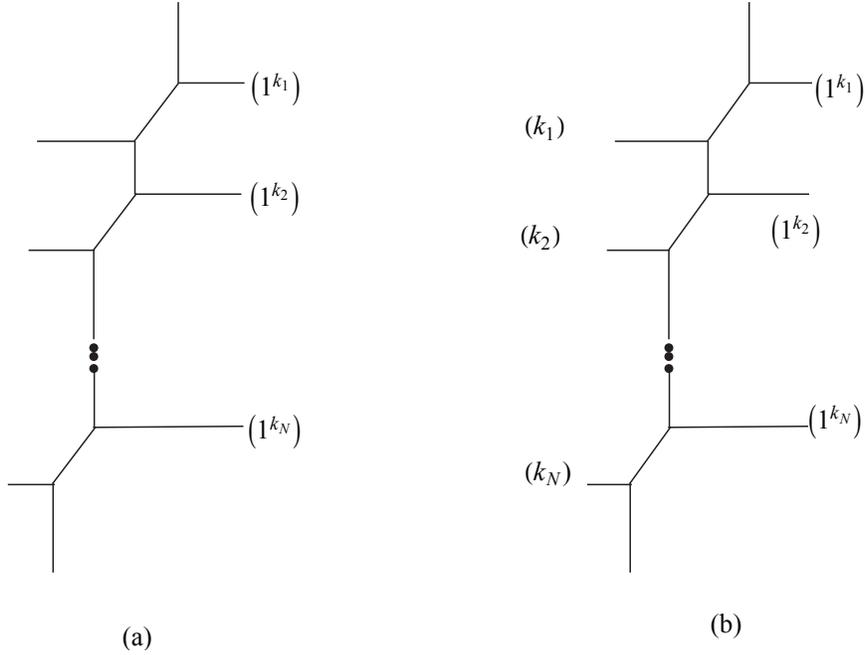}
\caption{Strip diagrams: (a) anti-fundamental, (b) adjoint}
\end{center}
\label{figure3}
\end{figure}

\subsection{Anti-fundamental matter}

In this subsection we compute the topological vertex on the strip with
boundary conditions given by single column Young diagram of variable lengths on one side of the strip
and
we show that there is a natural scaling limit on the Kahler moduli of the toric diagram amplitudes
such that these reduce to the vortex counting partition functions with anti-fundamentals.

We start from the (normalized) topological vertex on a strip as calculated in \cite{Iqbal}.
Its form and some properties useful to our computations are given in the appendix.

Let us compute then topological vertex on the strip with boundary conditions corresponding to
single columns representations on one side
and
trivial representations on the other.
It reads
$$
A_{\{\emptyset ,\emptyset ,\text{...},\emptyset \}}^{\left\{1^{k_1},1^{k_2},\text{...},1^{k_N}\right\}}=
\prod _{l=1}^N \prod_{i=1}^{k_l}\frac{1}{1-q^i}
\frac{
\prod_{l\leq m}^N \prod_{i=1}^{k_l}\left(1-Q_{\alpha_l\beta_m}q^{(i-1)}\right)
\prod_{l<m}^N\prod^{k_m}_{i=1}\left(1-Q_{\beta_l\alpha_m}q^{-(i-1)}\right)
}{
\prod_{l<m}^N \left(
\prod^{k_m}_{i=1}\left(1-Q_{\alpha_l\alpha_m} q^{i-1-k_l}\right)
\prod_{i=1}^{k_l}\left(1-Q_{\alpha_l\alpha_m} q^{1+k_m-i}\right)\right)
}.
$$
By defining
\begin{eqnarray}
Q_{\alpha_l\beta_f}&=&e^{-\beta(a_l+m_f)}\text{   }(l\leq f) \nonumber\\
Q_{\beta_f\alpha_m}&=&e^{\beta (a_l+m_f)}\text{   }(f<l)  \\
Q_{\alpha_l\alpha_m}&=& e^{\beta  a_{lm}}  \nonumber\\
q&=&e^{-\beta\hbar }  \nonumber
\end{eqnarray}
and
going to the cohomological limit $\beta\to0$ we find
$$
A_{\{\emptyset ,\emptyset ,\text{...},\emptyset \}}^{\left\{1^{k_1},1^{k_2},\text{...},1^{k_N}\right\}}
\to
\prod_{l=1}^N\prod_{i=1}^{k_l}\frac{1}{i \hbar }
\frac{
\prod_{l\leq f}^N \prod_{i=1}^{k_l}\left(a_l+m_f+(i-1)\hbar\right)
\prod_{f<l}^N \prod_{i=1}^{k_l}\left(a_l+m_f+(i-1)\hbar\right)
}{
\prod_{l<m}^N\prod_{i=1}^{k_m}\left(a_{ml}+\hbar \left(i-1-k_l\right)\right)
\prod_{i=1}^{k_l}\left(a_{lm}+\hbar \left(i-1-k_m\right)\right)
}
$$
which is easily recognized to be equal to (\ref{anti-vortex}).

\subsection{Adjoint matter}

As we said, the adjoint matter case can be obtained by computing the topological vertex on the strip
diagram of Fig.3b.

The topological vertex computation gives, by using the properties listed in the Appendix,
\begin{eqnarray}\label{adjvertex}
&&A_{\left\{k_1,k_2,\text{...},k_N\right\}}^{\left\{1^{k_1},1^{k_2},\text{...},1^{k_N}\right\}}=
\prod_{l=1}^N q^{k_l(k_l-1)/2}\prod_{i=1}^{k_l}\frac{1}{(1-q^i)^2}
\prod_{i=1}^{k_l}(1-q^iQ_{\alpha_l\beta_l})(1-q^{-i}Q_{\alpha_l\beta_l})
\times\\ &&
\frac{
\prod_{l<m}\prod_{i=1}^{k_l}(1-q^{i-1-k_m}Q_{\alpha_l\beta_m})(1-q^{i-1-k_m}Q_{\beta_l\alpha_m})
\prod_{i=1}^{k_m}(1-q^{-i+1+k_l}Q_{\alpha_l\beta_m})(1-q^{-i+1+k_l}Q_{\beta_l\alpha_m})}
{
\prod_{l<m}\prod_{i=1}^{k_l}(1-q^{i-1-k_m}Q_{\alpha_l\alpha_m})(1-q^{i-1-k_m}Q_{\beta_l\beta_m})
\prod_{i=1}^{k_m}(1-q^{-i+1+k_l}Q_{\alpha_l\alpha_m})(1-q^{-i+1+k_l}Q_{\beta_l\beta_m})}
\nonumber
\end{eqnarray}
where $\alpha_l=(1^{k_l})$ and $\beta_l=\alpha_l^t=(k_l)$.

Via the identifications
\begin{eqnarray}
q&=&e^{-\beta  \hbar }     \nonumber                 \\
Q_{\alpha _l\beta _l}&=&e^{-\beta  m}  \nonumber\\
\text{and for       } l&<&m  \nonumber\\
Q_{\beta_l\alpha_m}&=& e^{-\beta(m+a_{lm})}\nonumber\\
Q_{\alpha _l\alpha _m}&=&e^{-\beta  a_{lm} }\nonumber\\
Q_{\beta _l\beta _m}&=&e^{-\beta  a_{lm} }\nonumber\\
Q_{\alpha _l\beta _m}&=&e^{-\beta \left(a_{lm}-m\right)}\nonumber
\end{eqnarray}
and by taking the $\beta\longrightarrow0$ limit,  (\ref{adjvertex}) reduces to
\bea
\nonumber
&&\prod_{l=1}^N \prod_{i=1}^{k_l}\frac{(i\hbar+m)(i\hbar-m)}{(\hbar i)^2}
\prod_{l<m}^N
\frac{
\prod_{i=1}^{k_l}((i-1-k_m)\hbar +a_{lm}-m)((i-1-k_m)\hbar+a_{lm}+m)
}{
\prod_{i=1}^{k_l}((i-1-k_m)\hbar+a_{lm})((i-1-k_m)\hbar+a_{lm})} \\&&
\times \prod_{l<m}^N
\frac{
\prod_{i=1}^{k_m}((-i+1+k_l)\hbar+a_{lm}-m)((-i+1+k_l)\hbar +a_{lm}+m)}
{\prod_{i=1}^{k_m}((-i+1+k_l)\hbar+a_{lm})((-i+1+k_l)\hbar +a_{lm})}
\label{ciccio}
\eea
which is equal to
$$
Z^{adj}_{\bf k} ({\bf a},m)Z^{adj}_{\bf k}({\bf a},-m).
$$


\section{Surface operators and Toda CFT}

In this section we discuss the resummation formulae for supersymmetric vortex partition functions 
and interpret them in terms of suitable conformal blocks of Toda field theory.
In particular we provide a closed expression for the generating functions of vortices in terms 
of generalised hypergeometric functions, which in turn are the building blocks for amplitudes with
degenerate field insertions in Toda conformal field theory (CFT). As anticipated in the introduction
the origin of this relation has to be understood in terms of surface operators in four-dimensional
${\cal N}=2$ superconformal gauge theory, namely they can be described in terms of a two dimensional
gauge theory living on the defects where the surface operators lies.

In order to clarify this issue, 
let us consider the brane realization of surface operators 
in ${\cal N}=2$ SYM with $U(N)$ gauge group, see Fig.1.
The gauge theory is realized as a set of $N$ parallel D4-branes suspended between two parallel NS5 branes.
The transverse distance between these two NS5-branes is proportional to $\ln\Lambda$, $\Lambda$ being the
dynamical scale of the gauge theory \cite{four}.
The surface operator is obtained by suspending $N$ D2-branes between a further parallel and transversally displaced NS5'-brane
and the D4-branes. The transverse distance is the dynamical scale of a two dimensional theory, namely its Fayet-Iliopoulos parameter.
The location of the $N$ D2-branes on the D4-branes determines a partition of $N=\sum_{a=1}^N N_a$
corresponding to the generically unbroken gauge symmetry $\prod_a U(N_a)$.
We will consider the case of surface operators breaking to $U(1)^N$, namely $N_a=1$ for all $a$.
It was shown in \cite{bucov} that the abelian vortex partition function computes the classical limit of 
simple surface operators. In this section we argue that the non-abelian vortex counting of the previous sections
corresponds to the classical limit of interacting multiple surface operators of simple type.
Restricting to the computation of the classical value of the above surface operators
corresponds to move the two NS5-branes far away, therefore leaving the corresponding
$U(N)$ theory non dynamical. 
In particular the four dimensional gauge group becomes the flavour symmetry of the 
two dimensional gauge theory.

The gauge theory point of view also suggests looking for an AGT dual of the vortex partition function.
Actually, having realized the vortex partition function in terms of the dual topological string
as the vertex on the strip with single columns Young tableaux, we can formulate the Toda field theory dual
along the lines elaborated in \cite{KPW}, that is by realizing the surface operator insertions as particular toric branes
on the strip.

\begin{figure}
\begin{center}
\includegraphics[ width=0.7\textwidth]{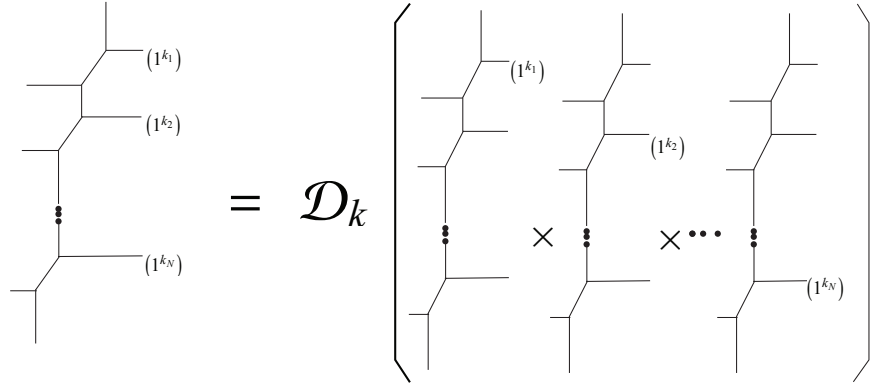}
\caption{The strip amplitude for matter in the anti-fundamental}
\end{center}
\label{figure4}
\end{figure}

The AGT dual of the Nekrasov partition function of the $U(N)$ gauge theory with $2N$ fundamentals can be obtained by 
the Toda conformal block on the sphere with two maximal punctures, at $0$ and $\infty$, and two semi-degenerate fields 
at $1$ and $z$ \cite{Wyllard}.
In this framework  the dual of surface operators is realized by inserting further degenerate fields \cite{surface} in the Toda field theory 
conformal block. Indeed we are about to prove that the resummed vortex partition function can be expressed precisely in terms of 
these conformal blocks.

Let us focus on the case of antifundamental matter and consider the following generating function 
\be
{\cal Z}^{af}({\bf z}, m_f, a_l, \hbar)=
\sum_{\tt k} {\bf z}^{\tt k} Z^{af}_{\tt k}
\label{zeta}\ee
where ${\tt k}=\{k_1,\ldots,k_N\}$, ${\bf z}=\{z_1,\ldots,z_N\}$ and ${\bf z}^{\tt k}=\prod_l z_l^{k_l}$
By making use of the identity
\begin{eqnarray}
(a-l)_m(-a-m)_l=\left(1+\frac{(m-l)}{a}\right)^{-1}(a+1)_m(-a+1)_l
\label{identity}
\end{eqnarray}
where $(a)_n=\prod _{i=1}^n(a+i-1)$ is the usual Pochhammer symbol
we can rewrite the vortex partition function as
\begin{eqnarray}
 Z^{af}_{\tt k}
=
\underset{l<m}{\overset{N}{\prod }}\left(1+ \hbar\frac{(k_m-k_l)}{a_{m l}}\right)\prod_{l=1}^N 
\frac{1}{k_l!}\prod _f^N\left(\frac{a_l+m_f}{\hbar }\right)_{k_l}
\left(
\underset{l\neq m}{\overset{N}{\prod }}
\left(\frac{a_{l m}+\hbar}{\hbar }\right)_{k_l}\right)^{-1}
\end{eqnarray}
By replacing the latter in the definition (\ref{zeta}), we then get
\begin{eqnarray}
{\cal Z}^{af}({\bf z}, m_f, a_l, \hbar)=
D \prod _{l=1}^N\, _N\, F_{N-1}\left(A_l,B_l,z_l\right)
\label{zaff}
\end{eqnarray}
where $\, _N\, F_{N-1}\left(A,B,z\right)
=
\sum_k \frac{z^k}{k!}\frac{(A_1)_k\cdot\ldots\cdot(A_N)_k}{(B_1)_k\cdot\ldots\cdot(B_{N-1})_k}
$  is the generalized hypergeometric function and 
\begin{eqnarray}
\label{D}
D&=&\underset{l<m}{\overset{N}{\prod }}\left(1+\hbar\frac{z_m\partial _{z_m}-z_l\partial _{z_l}}{a_{m l}}\right)  \\
A_l&=&\left\{\frac{a_l+m_1}{\hbar },\frac{a_l+m_2}{\hbar },\text{...},\frac{a_l+m_N}{\hbar }\right\}   \nonumber \\
B_l&=&\left\{\frac{a_{l 1}+\hbar}{\hbar },\frac{a_{l 2}+\hbar}{\hbar },\text{...},\frac{a_{l N}+\hbar}{\hbar }\right\}  \nonumber
\end{eqnarray}
The AGT dual picture is then recovered by noticing that the generalized hypergeometric functions are 
the degenerate conformal blocks in Toda field theory considered in \cite{FL}, namely the
ones associated to the four point function 
\be
<\alpha_2|V_{-b{\omega_1}}(z)V_{-\kappa{\omega_{N-1}}}(1)|\alpha_1>
\ee
where $|\alpha_1>$ and $|\alpha_2>$ are two primary states, $V_{-b{\omega_1}}$ is the highest weight degenerate field
and $V_{-\kappa{\omega_{N-1}}}$ the vertex with momentum proportional to the lowest root.
Each of them corresponds to the field theory limit of a single toric brane amplitude \cite{KPW}.
The total amplitude (\ref{zaff}) is 
given by the action of the differential operator $D$ in (\ref{D}) over a product of 
$N$ single brane amplitudes (see fig. 4). The non-abelian structure of the amplitude is encoded in the operator $D$
of which it would be nice to provide a precise CFT transliteration. 

As we have shown in section 2.2, the vortex counting can be obtained from instanton counting by restricting to columns.
This should have a clean counterpart in the AGT dual picture.
Notice that the full amplitude is expressed in terms of correlators with a single degenerate field insertion.
Therefore it should be possible to interpret (\ref{zaff}) as a correlator on a degenerate sphere, with further insertions of degenerate fields 
on the stretching collars. In this way, the intermediate states would reduce to a tower of degenerate states
which depend on the level only and thus could be represented as columns with height corresponding to the level.

Let us notice that the operator 
$z\partial_z$ acting on generalized hypergeometric functions 
produces linear combinations of them with shifted parameters. Therefore 
formula (\ref{zaff}) can also be written in terms of products of linear combinations of generalized hypergeometric
functions with shifted parameters.

It is easy to uplift the previous procedure to the full open topological string amplitude on the strip
\begin{eqnarray}
A_{\{\emptyset ,\emptyset ,\text{...},\emptyset \}}^{\left\{1^{k_1},1^{k_2},\text{...},1^{k_N}\right\}}&=&\prod _{l<m}^N\frac{1-Q_{\alpha _l\alpha _m}q^{k_l-k_m}}{1-Q_{\alpha _a\alpha _b}}\left(\frac{Q_{\beta _l\alpha _m}q}{Q_{\alpha _l\alpha _m}}\right)_{k_m}  \\
&&\prod _{l=1}^N \underset{i=1}{\overset{k_l}{\prod }}\frac{1-Q_{\alpha _l\beta _l}q^{i-1}}{1-q^i}\overset{N}{\prod _{l<m} }\overset{k_l}{\prod _{i=1} }\frac{1-Q_{\alpha _l\beta _m}q^{i-1}}{1-Q_{\alpha _l\alpha _m} q^i}\overset{k_m}{\prod _{i=1} }\frac{1-Q^{-1}{}_{\beta _l\alpha _m} q^{i-1}}{1-Q^{-1}{}_{\alpha _l\alpha _m} q^i} \nonumber
\end{eqnarray}
For $l<m$, we define $ M_{l,m}=Q_{\alpha _l\beta _m}q^{-1}; M_{m,l}=Q^{-1}{}_{\beta _l\alpha _m}q^{-1};Q_{l,m}=Q_{\alpha _l\alpha _m};Q_{m,l}=Q^{-1}{}_{\alpha _l\alpha _m}$,while for $l=m$, $M_{l,l}=Q_{\alpha _l\beta _l}q^{-1}; Q_{l,l}=1$. By also defining
\begin{eqnarray}
[Q]_k&\text{=}&\prod _{i=1}^k\left(1-Q q^i\right)  \\
\mathcal{D}_{{\tt k}}&\text{=}&\prod _{l<m}^N\frac{1-Q_{l,m}q^{k_l-k_m}}{1-Q_{l,m}}\left(\frac{M_{m,l}^{-1}}{Q_{l,m}}\right)_{k_m}  \nonumber
\end{eqnarray}
we get
\be
A_{\{\emptyset ,\emptyset ,\text{...},\emptyset \}}^{\left\{1^{k_1},1^{k_2},\text{...},1^{k_N}\right\}}=
\prod _{m=1}^N 
\mathcal{D}_{{\tt k}}
\frac{\prod _{l=1}^N\left[M_{l,m}\right]_{k_l}}{[1]_{k_m}\prod _{n\neq m}^N\left[Q_{n,m}\right]_{k_n}}.
\ee
This is schematically encoded in Fig. 4.
By resumming the topological string amplitudes as
\be
{\cal A}({\bf z})=\sum_{{\tt k}}{\bf z}^{\tt k}
A_{\{\emptyset ,\emptyset ,\text{...},\emptyset \}}^{\left\{1^{k_1},1^{k_2},\text{...},1^{k_N}\right\}}
\ee
we obtain 
\be
{\cal A}({\bf z})={\cal D}\prod_{l=1}^N {}_N\, \Phi_{N-1}\left(X_l,Y_l,z_l\right)
\ee
where $ {}_N\Phi_{N-1}\left(X,Y,z\right)=
\sum_k \frac{z^k}{[1]_k}\frac{[X_1]_k\cdot\ldots\cdot[X_N]_k}{[Y_1]_k\cdot\ldots\cdot[Y_{N-1}]_k}
$
is a q-deformed generalized hypergeometric function,
$X_l=e^{-\beta\hbar(A_l-1)}$,
$Y_l=e^{-\beta\hbar B_l}$
and 
${\cal D}=\prod_{l<m}\frac{1-Q_{l,m}q^{z_l\partial_{z_l}-z_m\partial_{z_m}}}{1-Q_{l,m}}$
up to a multiplicative redefinition of the open moduli ${\bf z}$.
The operator ${\cal D}$ is a finite difference operator whose action on the
q-deformed generalized hypergeometric functions multiplicatively shifts their arguments.
This result could be interpreted in the light of a five dimensional uplift
of the AGT relation \cite{qdef}.


Let us now discuss the vortex partition function for the adjoint matter case.
By making use of the previous identity (\ref{identity})
we obtain
\be
Z^{adj}_{{\tt k}}=
\prod_{l<m}\frac{\left(1-\hbar\frac{k_l-k_m}{a_{lm}}\right)}{\left(1-\hbar\frac{k_l-k_m}{a_{lm}-m}\right)}
\prod_l \frac{(m/\hbar+1)_{k_l}}{k_l!}
\prod_{l\ne m}\frac{\left(\frac{a_{lm}-m}{\hbar}+1\right)_{k_l}}
{\left(\frac{a_{lm}}{\hbar}+1\right)_{k_l}}
\prod_{l<m}
\frac{\left(-\frac{a_{lm}+m}{\hbar}-k_l\right)_{k_m}}
{\left(-\frac{a_{lm}-m}{\hbar}-k_l\right)_{k_m}}
\label{paperino}
\ee
Notice that this form does not show an obvious resummation in terms of generalized hypergeometric 
functions due to the last multiplicative factor in (\ref{paperino}).
However, the open topological string amplitude in the $\beta\to 0$ limit (\ref{ciccio})
can be recast, by making use of (\ref{paperino}), in the form\footnote{Notice that in the product the 
two multiplicative unfair terms cancel.}
\be
\prod_{l<m}\frac{\left(1-\hbar\frac{k_l-k_m}{a_{lm}}\right)^2}{\left(1-\hbar\frac{k_l-k_m}{a_{lm}-m}\right)
\left(1-\hbar\frac{k_l-k_m}{a_{lm}+m}\right)}
\prod_{l,m}\frac{\left(\frac{a_{lm}-m}{\hbar}+1\right)_{k_l}
\left(\frac{a_{lm}+m}{\hbar}+1\right)_{k_l}}
{\left(\left(\frac{a_{lm}}{\hbar}+1\right)_{k_l}\right)^2}
\label{brutta}
\ee
By resumming the above coefficients against ${\bf z}^{\tt k}$
one finally gets
\be
{\cal D}^{adj}({\bf a},m)\prod_l {}_{2N}F_{2N-1}\left(A^{adj}_l,B^{adj}_l,z_l\right)
\label{paperone}
\ee
where
\bea
A^{adj}_l&=&\left(\frac{a_{lm}+m}{\hbar}+1,\frac{a_{lm}-m}{\hbar}+1\right)\nonumber\\
B^{adj}_l&=&\left(\frac{a_{lm}}{\hbar}+1,\frac{a_{lm}}{\hbar}+1\right)
\eea
and
\be
{\cal D}^{adj}({\bf a},m)=
\prod_{l<m}\frac{\left(1-\hbar\frac{z_l\partial_{z_l}-z_m\partial_{z_m}}{a_{lm}}\right)^2}{\left(1-\hbar\frac
{z_l\partial_{z_l}-z_m\partial_{z_m}}
{a_{lm}-m}\right)
\left(1-\hbar\frac
{z_l\partial_{z_l}-z_m\partial_{z_m}}
{a_{lm}+m}\right)}.
\ee
The resummed form (\ref{paperone}) in terms of generalized hypergeometric functions
suggests an interpretation of the resummed open topological string amplitude in the $\beta\to0$ limit
as degenerate conformal blocks of Toda field theory on the sphere.
We argue that, by using a suitable generalization of the results in \cite{flo}
to Toda field theory, this can be recast as conformal blocks on the torus
giving the expected AGT dual description.

As it is well known generalized hypergeometric functions satisfy generalized hypergeometric 
{\it differential} equations. Moreover, the q-deformed generalized hypergeometric functions,
resumming the vertex amplitudes, satisfy corresponding {\it finite difference} equations.

\section{Discussion and Open Issues}

In this paper we presented a description of the moduli spaces of non-abelian $U(N)$ vortices
with adjoint and $N$ (anti-)fundamental matter multiplets as holomorphic submanifolds of instanton
moduli spaces. The associated partition functions provide the classical limit (zero instanton sector)
of the v.e.v. of  multiple surface operator insertions in the parent ${\cal N}=2$ superconformal gauge theories
in four dimensions.
The results we found can be simply expressed in terms of an ensemble of abelian partition functions 
intertwined by the action of a differential operator which couples the abelian factors of the Cartan subgroup,
and thus induces pairwise interactions in the ensemble of multiple surface operators of simple type.

We performed a resummation of the full partition functions
over the vortex numbers by providing a closed expression in terms of combinations of generalized hypergeometric functions.
This allowed us to make contact with a dual Toda CFT description in terms
conformal blocks with degenerate field insertions. In particular we have shown that the vortex counting amounts to a 
restriction of instanton counting just to column diagrams and proposed a possible interpretation in the CFT dual which
should be further refined. 

We also studied the K-theoretical uplift of these countings and find a dual string description in terms of open
topological strings on a strip with suitable boundary conditions.

There are several issues raised by our results which are worth to be investigated further.
First of all it would be highly desirable to provide a full four-dimensional computation of the
instanton partition function with interacting surface operators, going beyond the classical limit
presented in this paper. It would be also interesting to investigate the extension and the relation 
to other kind of surface operators, for example to full ones \cite{full}.
Concerning the K-theoretical uplift, a nice connection of the abelian vortex counting with 
the equivariant $J$-function \cite{givental} encoding the quantum cohomology of complex projective spaces
has been pointed out in \cite{bucov}. It is natural to argue that the generating functions we find
in this paper are related to $J$-functions of more general flag varieties.
Along this line of thought, it would be certainly interesting to analyse the moduli space of vortices
on generic Riemann surfaces in order to extend these relations to equivariant Gromov-Witten
invariants of higher genera. A useful starting point should be \cite{baptista}
and the analysis of \cite{biswas}.
 
A complementary route that could be taken in this direction is to analyse the B-model mirror
description of the strip computations that we presented. Indeed we showed that the resummation of vortices
can be performed also at the K-theoretical level in terms of q-deformed generalised hypergeometric functions,
which point to the possibility of encoding in geometrical terms the fully resummed amplitudes.
The route to the B-model mirror picture could pass by a rephrasing of the result in terms of generalized matrix models
\cite{mm} via the encoding of the mirror geometry in the spectral curve.
All this points to an heavy role played by integrable systems also in vortex counting problems, both from their 
appearance in the AGT dual \cite{hitchin} and from the topological string viewpoint \cite{tsih,braini}.
Furthermore, in \cite{rubik}, it has been shown that the approach of \cite{NS} can be recast in terms of 
restriction of the instanton counting to columns diagrams. It would be nice to exploit this observation to make a precise 
connection between vortex counting and the Nekrasov-Shatashvili limit.

It would be nice to further analyse the role of vortex counting in the AGT correspondence
also in the light of the application to fractional quantum Hall systems presented in \cite{ST}.

Last but not least, it has been shown in \cite{fmpt} that the instanton counting techniques are suitable
to describe the superpotentials of ${\cal N}=1$ theories in four dimensions by setting
the Cartan parameters to appropriate values, describing the ${\cal N}=1$ vacua. An evidence in this direction 
is that in \cite{akv,capo,braini} disk amplitudes are expressed precisely in terms of hypergeometric functions
with parameters fixed in terms of the masses and the strong coupling scale.
We expect that analogous results can be obtained in the vortex counting case, possibly opening a window on
a extension of AGT duality to ${\cal N}=1$ theories.

\vspace{2cm}

{\bf Acknowledgements}
We would like to thank
F.~Benini, A.~Brini, K.~Maruyoshi, S.~Pasquetti and R.~Poghossian
for useful discussions.
Z.J. thanks H.~Liu for discussions.
G.B. and Z.J. are partially supported by the INFN project TV12. 
A.T. is partially supported by  PRIN ``Geometria delle variet\`a algebriche e loro spazi di moduli'' 
and the INFN project PI14 ``Nonperturbative dynamics of gauge theories''.

\appendix

\section*{Appendix}

\section{The conventions on the topological vertex}

In this Appendix we summarize the usual conventions on the topological vertex on the strip
and some useful formulas that we used in the main text.

The normalized amplitude on the strip is given by \cite{Iqbal} 
\begin{eqnarray}
A_{\{\beta \}}^{\{\alpha \}}&=&\prod _{a=1}^N s_{\alpha _a}s_{\beta _a}\overset{\infty }{\prod _{i=-\infty } }\prod _{a\leq b} \left(1-q^iQ_{\alpha _a\beta _b}\right){}^{C_i\left(\alpha _a,\beta _b\right)}\prod _{a<b} \left(1-q^iQ_{\beta _a\alpha _b}\right){}^{C_i\left(\beta _a^t,\alpha _b^t\right)}\\
&&\left(\prod _{a<b} \left(1-q^iQ_{\alpha _a\alpha _b}\right){}^{C_i\left(\alpha _a,\alpha _b^t\right)}\left(1-q^iQ_{\beta _a\beta _b}\right){}^{C_i\left(\beta _a^t,\beta _b\right)}\right){}^{-1}\nonumber
\end{eqnarray}
where $\alpha_a,\beta_b$ are the left and right partitions
parametrizing the toric branes boundary conditions.
$s_{\alpha}$ is the Schur function
\begin{eqnarray}
s_{\alpha }(q)\text{=}q^{\sum _i(i-1)\alpha _i}\prod _{p\in \alpha } \frac{1}{1-q^{\text{hook}(p)}}
\end{eqnarray}
where $\alpha_i$ is the $i-th$ component of the partition $\alpha$, and $hook(p)$ is the hook length
of a point $p\in \alpha$ seen as a Young tableaux.

For columns and strips one has
\begin{eqnarray}
s_{\left(1^k\right)}&=&\prod _{i=1}^k \frac{1}{1-q^i}   \nonumber\\
s_{(k)}&=&q^{\frac{k(k-1)}{2}}\prod _{i=1}^k \frac{1}{1-q^i}  
\end{eqnarray}

The coefficients
$C_k(\alpha,\beta)$ are defined for two given partitions $\alpha$ and $\beta$ by the formula
\begin{eqnarray}
\sum _k C_k(\alpha ,\beta )q^k&=&\frac{q}{(1-q)^2}\left(1+(q-1)^2\sum _{i=1}^{d_{\alpha }} q^{-i}\sum _{j=0}^{\alpha _i-1} q^j\right)\nonumber\\
&&\left(1+(q-1)^2\sum _{i=1}^{d_{\beta }} q^{-i}\sum _{j=0}^{\beta _i-1} q^j\right)-\frac{q}{(1-q)^2}
\end{eqnarray}
and are symmetric by definitions, that is $C_i(\alpha ,\beta )=C_i(\beta ,\alpha )$.

Specializing to columns and strips one finds
\begin{eqnarray}
C_i\left(1^k,\emptyset \right)&=&\left\{
\begin{array}{cc}
 1 & i\in [0,k-1]\nonumber \\
 0 & \text{otherwise}
\end{array}
\right. \\
C_i((k),\emptyset )&=&\left\{
\begin{array}{cc}
 1 & i\in [-k+1,0] \nonumber\\
 0 & \text{otherwise}
\end{array}
\right. \\
C_i\left(1^{k_1},(k_2)\right)&=&\left\{
\begin{array}{cc}
 1 & i\in \left[-k_2,k_1-k_2-1\right]\cup \left[k_1-k_2+1,k_1\right] \nonumber\\
 0 & \text{otherwise}
\end{array}
\right.
\end{eqnarray}


\begin{thebibliography}{99}
\bibitem{akv}
M.~Aganagic, A.~Klemm and C.~Vafa,
  Z.\ Naturforsch.\  A {\bf 57} (2002) 1
  [arXiv:hep-th/0105045].

\bibitem{tv}
M.~Aganagic, A.~Klemm, M.~Marino and C.~Vafa,
  Commun.\ Math.\ Phys.\  {\bf 254}, 425 (2005)
  [arXiv:hep-th/0305132].

\bibitem{tsih}
M.~Aganagic, R.~Dijkgraaf, A.~Klemm, M.~Marino and C.~Vafa,
  Commun.\ Math.\ Phys.\  {\bf 261} (2006) 451
  [arXiv:hep-th/0312085].

\bibitem{AGT}
L.~F.~Alday, D.~Gaiotto and Y.~Tachikawa,
  ``Liouville Correlation Functions from Four-dimensional Gauge Theories,''
  arXiv:0906.3219 [hep-th].

\bibitem{surface}
L.~F.~Alday, D.~Gaiotto, S.~Gukov, Y.~Tachikawa and H.~Verlinde,
  ``Loop and surface operators in N=2 gauge theory and Liouville modular geometry,''
  arXiv:0909.0945 [hep-th].

\bibitem{full}
L.~F.~Alday and Y.~Tachikawa,
  Lett.\ Math.\ Phys.\  {\bf 94} (2010) 87
  [arXiv:1005.4469 [hep-th]].
 A.~Braverman, B.~Feigin, M.~Finkelberg and L.~Rybnikov,
  arXiv:1008.3655 [math.AG].

\bibitem{qdef}
 H.~Awata and Y.~Yamada,
  JHEP {\bf 1001} (2010) 125
  [arXiv:0910.4431 [hep-th]].

\bibitem{baptista}
J.~M.~Baptista,
  Commun.\ Math.\ Phys.\  {\bf 291} (2009) 799
  [arXiv:0810.3220 [hep-th]].

\bibitem{BPZ}
A.~A.~Belavin, A.~M.~Polyakov and A.~B.~Zamolodchikov,
  Nucl.\ Phys.\  B {\bf 241} (1984) 333.

\bibitem{biswas}
I.~Biswas, N.~M.~Romao,
``Moduli of vortices and Grassmann manifolds, ``
arXiv:1012.4023 [math].

\bibitem{braini}
A.~Brini, ``Open topological strings and integrable hierarchies: Remodeling the A-model'', to appear.

\bibitem{also}
A.~Brini, M.~Marino and S.~Stevan,
  arXiv:1010.1210 [hep-th].
A.~Marshakov, A.~Mironov and A.~Morozov,
  Teor.\ Mat.\ Fiz.\  {\bf 164} (2010) 1:3
  [arXiv:1011.4491 [hep-th]].
 M.~Taki,
  arXiv:1007.2524 [hep-th].
D.~Gaiotto,
  arXiv:0911.1316 [hep-th].
H.~Awata, H.~Fuji, H.~Kanno, M.~Manabe and Y.~Yamada,
  ``Localization with a Surface Operator, Irregular Conformal Blocks and Open
  Topological String,''
  arXiv:1008.0574 [hep-th].
K.~Maruyoshi and M.~Taki,
  Nucl.\ Phys.\  B {\bf 841}, 388 (2010)
  [arXiv:1006.4505 [hep-th]].
U.~Bruzzo, W.~y.~Chuang, D.~E.~Diaconescu, M.~Jardim, G.~Pan and Y.~Zhang,
  ``D-branes, surface operators, and ADHM quiver representations,''
  arXiv:1012.1826 [hep-th].
 A.~Mironov and A.~Morozov,
  J.\ Phys.\ A  {\bf 43} (2010) 195401
  [arXiv:0911.2396 [hep-th]].
 A.~Mironov and A.~Morozov,
  JHEP {\bf 1004} (2010) 040
  [arXiv:0910.5670 [hep-th]].


\bibitem{hitchin}
G.~Bonelli and A.~Tanzini,
  Phys.\ Lett.\  B {\bf 691} (2010) 111
  [arXiv:0909.4031 [hep-th]].
J.~Teschner,
  arXiv:1005.2846 [hep-th].

\bibitem{tanzinietc}
U.~Bruzzo, F.~Fucito, J.~F.Morales, A.~Tanzini,
  JHEP {\bf 0305}, 054 (2003)
  [arXiv:hep-th/0211108].

\bibitem{capo}
N.~Caporaso, L.~Griguolo, M.~Marino, S.~Pasquetti and D.~Seminara,
 Phys.\ Rev.\  D {\bf 75}, 046004 (2007)
 [arXiv:hep-th/0606120].
 
\bibitem{mm}
M.~C.~N.~Cheng, R.~Dijkgraaf and C.~Vafa,
  arXiv:1010.4573 [hep-th].
  R.~Dijkgraaf and C.~Vafa,
  arXiv:0909.2453 [hep-th].
K.~Maruyoshi and F.~Yagi,
  JHEP {\bf 1101} (2011) 042
  [arXiv:1009.5553 [hep-th]].
G.~Bonelli, K.~Maruyoshi, A.~Tanzini and F.~Yagi,
  arXiv:1011.5417 [hep-th].

\bibitem{bucov}
T.~Dimofte, S.~Gukov and L.~Hollands,
  arXiv:1006.0977 [hep-th].

\bibitem{eto1}
M.~Eto, Y.~Isozumi, M.~Nitta, K.~Ohashi and N.~Sakai,
  Phys.\ Rev.\ Lett.\  {\bf 96} (2006) 161601
  [arXiv:hep-th/0511088].

\bibitem{eto2}
M.~Eto, Y.~Isozumi, M.~Nitta, K.~Ohashi and N.~Sakai,
  J.\ Phys.\ A  {\bf 39} (2006) R315
  [arXiv:hep-th/0602170].

\bibitem{FL}
V.~A.~Fateev and A.~V.~Litvinov,
  ``Correlation functions in conformal Toda field theory I,''
  JHEP {\bf 0711} (2007) 002
  [arXiv:0709.3806 [hep-th]].


\bibitem{flo}
V.~A.~Fateev, A.~V.~Litvinov, A.~Neveu and E.~Onofri,
  J.\ Phys.\ A  {\bf 42} (2009) 304011
  [arXiv:0902.1331 [hep-th]].

\bibitem{PF}
 R.~Flume and R.~Poghossian,
  Int.\ J.\ Mod.\ Phys.\  A {\bf 18}, 2541 (2003)
  [arXiv:hep-th/0208176].
  
\bibitem{fmpt}
F.~Fucito, J.~F.~Morales, R.~Poghossian and A.~Tanzini,
 JHEP {\bf 0601}, 031 (2006)
 [arXiv:hep-th/0510173].

\bibitem{N=2}
D.~Gaiotto,
  ``N=2 dualities,''
  arXiv:0904.2715 [hep-th].

\bibitem{GL}
A.~A.~Gerasimov and D.~R.~Lebedev,
  ``On topological field theory representation of higher analogs of classical
  special functions,''
  arXiv:1011.0403 [hep-th].

\bibitem{givental}
A.~Givental,
``A mirror theorem for toric complete intersections'',
in ``Topological field theory, Primitive forms and related topics''
(Kyoto, 1996), Progr. Math {\bf 160} (1998), 141-175,
[arXiv:alg-geom/9701016];
A.~Givental, Y.-P.~ Lee,
``Quantum K-theory on flag manifolds, finite-difference Toda lattices and quantum groups'',
arXiv:math/0108105.

\bibitem{hh}
A.~Hanany and K.~Hori,
  Nucl.\ Phys.\  B {\bf 513} (1998) 119
  [arXiv:hep-th/9707192].


\bibitem{ht}
  A.~Hanany and D.~Tong,
  JHEP {\bf 0307} (2003) 037
  [arXiv:hep-th/0306150].

\bibitem{shifman}
R. Auzzi, S. Bolognesi, J. Evslin, K. Konishi and A. Yung,
Nucl.\ Phys.\ B {\bf 673}, 187 (2003) [hep-th/0307287];
M. Shifman and A. Yung,
Phys.\ Rev.\ D {\bf 70}, 045004 (2004) [hep-th/0403149];
A. Hanany and D. Tong,
JHEP {\bf 0404}, 066 (2004) [hep-th/0403158].

\bibitem{giappi}
T.~Fujimori, T.~Kimura, M.~Nitta and K.~Ohashi,
  ``Vortex counting from field theory,''
  arXiv:1204.1968 [hep-th].

\bibitem{Iqbal}
 A.~Iqbal and A.~K.~Kashani-Poor,
  Adv.\ Theor.\ Math.\ Phys.\  {\bf 10} (2006) 317
  [arXiv:hep-th/0410174].


 \bibitem{vafa}
S.~H.~Katz, A.~Klemm and C.~Vafa,
  Nucl.\ Phys.\  B {\bf 497} (1997) 173
  [arXiv:hep-th/9609239].

\bibitem{KPW}
C.~Kozcaz, S.~Pasquetti and N.~Wyllard,
  JHEP {\bf 1008} (2010) 042
  [arXiv:1004.2025 [hep-th]].




\bibitem{russi}
 A.~Mironov and A.~Morozov,
  Phys.\ Lett.\  B {\bf 680}, 188 (2009)
  [arXiv:0908.2190 [hep-th]].
A.~Mironov and A.~Morozov,
  Nucl.\ Phys.\  B {\bf 825}, 1 (2010)
  [arXiv:0908.2569 [hep-th]].


\bibitem{Nek}
N.~A.~Nekrasov,
  Adv.\ Theor.\ Math.\ Phys.\  {\bf 7}, 831 (2004)
  [arXiv:hep-th/0206161]


\bibitem{NS}
N.~A.~Nekrasov and S.~L.~Shatashvili,
  ``Quantization of Integrable Systems and Four Dimensional Gauge Theories,''
  arXiv:0908.4052 [hep-th].

\bibitem{rubik}
R.~Poghossian,
  ``Deforming SW curve,''
  arXiv:1006.4822 [hep-th].


\bibitem{ST}
R.~Santachiara and A.~Tanzini,
  Phys.\ Rev.\  D {\bf 82}, 126006 (2010)
  [arXiv:1002.5017 [hep-th]].




\bibitem{shadchin}
S.~Shadchin,
  JHEP {\bf 0708}, 052 (2007)
  [arXiv:hep-th/0611278].


\bibitem{topos}
E.~Witten,
  Commun.\ Math.\ Phys.\  {\bf 118}, 411 (1988).



\bibitem{four}
E.~Witten,
  ``Solutions of four-dimensional field theories via M-theory,''
  Nucl.\ Phys.\  B {\bf 500} (1997) 3
  [arXiv:hep-th/9703166].

\bibitem{wittensurf}
E.~Witten,
  Fortsch.\ Phys.\  {\bf 55}, 545 (2007).
S.~Gukov and E.~Witten,
  arXiv:0804.1561 [hep-th].

\bibitem{Wyllard}
N.~Wyllard,
  JHEP {\bf 0911}, 002 (2009)
  [arXiv:0907.2189 [hep-th]].



\bibitem{Yoshida}
Y.~Yoshida,
  ``Localization of Vortex Partition Functions in $\mathcal{N}=(2,2) $ Super
  Yang-Mills theory,''
  arXiv:1101.0872 [hep-th].

\end{thebibliography}

\end{document}